\documentclass[aip, jcp, reprint, article]{revtex4-1}

\usepackage{amsmath,amsfonts,amssymb}						% the holy trinity
\usepackage{braket} 										% bras and kets
\usepackage{nicefrac} 										% smaller fractions
\usepackage{mhsetup} 										% for mathtools
\usepackage{mathtools}										% tweaking math appearance
\usepackage[version=3]{mhchem} 								% chemical formulas
\usepackage{siunitx} 										% units

\DeclareSIUnit\kay{\per \cm}

\newcommand*{\tabref}[1]{\tablename~\ref{#1}}
\newcommand*{\figref}[1]{\figurename~\ref{#1}}
\newcommand*{\equref}[1]{Eq.~\eqref{#1}}

\newcommand{\bands}[2]{\mbox{(#1--#2)}}		% mbox is for avoiding line breaks.
\newcommand{\fone}{$\mathrm{F}_1$ }
\newcommand{\ftwo}{$\mathrm{F}_2$}
\newcommand{\Xstate}{$1^2\Sigma^+$~}
\newcommand{\bstate}{$2^2\Sigma^+$~}
\newcommand{\astate}{$1^2\Pi$~}
\newcommand{\LiSr}{\ce{^7Li^{88}Sr}~}

\newcommand*{\citeText}[1]{%	cite without footnote
	\begingroup
	\romannumeral-`\x % remove space at the beginning of \setcitestyle
	\setcitestyle{numbers}%
	\cite{#1}%
	\endgroup   
}

\begin{document}

\title[\LiSr \Xstate and \bstate states]{Laser and Fourier transform spectroscopy of \LiSr}

\date{\today}

\begin{abstract}
	This is an author-created, un-copyedited version of an article published in the Journal of Physics B. IOP Publishing Ltd is not responsible for any errors or omissions in this version of the manuscript or any version derived from it. The Version of Record is available online at https://doi.org/10.1088/1361-6455/aa8ca0.
	
	\begin{center}
	------------------------------------------------------------------
	\end{center}
	
	\ce{LiSr} was produced in a heat-pipe oven and its thermal emission spectrum around \SI{9300}{\kay} was recorded by a high resolution Fourier transform spectrometer. In addition, selected lines of the spectrum of deeply bound vibrational levels of the \Xstate and \bstate states were studied using laser excitation to facilitate the assignment of the lines. The ground state could be described for $v^{\prime\prime} =$ \numrange{0}{2}, $N^{\prime\prime}$ up to 105 and the \bstate state for $v^{\prime} = 0$ up to $N^\prime = 68$. For both states, Dunham coefficients, spin-rotation parameters and potential energy curves were evaluated. A coupling of the \bstate state to the \astate state was observed, allowing a local description with Dunham coefficients of the \astate state and an approximate evaluation of the coupling strength.
\end{abstract}

\author{Erik Schwanke}
\affiliation{ Leibniz Universit\"at Hannover, Institute of Quantum Optics, Welfengarten 1, 30167 Hannover, Germany}
\affiliation{Leibniz Universit\"at Hannover, Laboratory for Nano- and Quantum Engineering,  Schneiderberg 39, 30167 Hannover, Germany}
\email{schwanke@iqo.uni-hannover.de}

\author{Horst Kn\"ockel}
\affiliation{ Leibniz Universit\"at Hannover, Institute of Quantum Optics, Welfengarten 1, 30167 Hannover, Germany}
\affiliation{Leibniz Universit\"at Hannover, Laboratory for Nano- and Quantum Engineering,  Schneiderberg 39, 30167 Hannover, Germany}

\author{Asen Pashov}
\affiliation{ Department of Physics, Sofia University, 5 James Bourchier Boulevard, 1164 Sofia, Bulgaria}

\author{Alexander Stein}
\affiliation{ Leibniz Universit\"at Hannover, Institute of Quantum Optics, Welfengarten 1, 30167 Hannover, Germany}

\author{Silke Ospelkaus}
\affiliation{ Leibniz Universit\"at Hannover, Institute of Quantum Optics, Welfengarten 1, 30167 Hannover, Germany}
\affiliation{Leibniz Universit\"at Hannover, Laboratory for Nano- and Quantum Engineering,  Schneiderberg 39, 30167 Hannover, Germany}

\author{Eberhard Tiemann}
\affiliation{ Leibniz Universit\"at Hannover, Institute of Quantum Optics, Welfengarten 1, 30167 Hannover, Germany}
\affiliation{Leibniz Universit\"at Hannover, Laboratory for Nano- and Quantum Engineering,  Schneiderberg 39, 30167 Hannover, Germany}

\keywords{PACS 31.50.-x 	Potential energy surfaces ,
	PACS 33.15.Mt  	Rotation, vibration, and vibration-rotation constants, 
	PACS 33.20.-t 	Molecular spectra}

\maketitle
	
\section*{Introduction}

	Molecules consisting of one alkali-metal atom and one alkaline-earth atom receive rising interest for their prospective value in the field of ultracold quantum gases because their ground state  \Xstate with its electric and a magnetic dipole moment offers advantageous properties (see e.g. references \citeText{tscherbul_controlling_2006,pasquiou_quantum_2013,roy_photoassociative_2016}).
	\ce{RbSr}, \ce{RbYb} and \ce{LiYb} have been studied in weakly bound states via Feshbach-resonance spectroscopy (see references \citeText{pasquiou_quantum_2013,bruni_observation_2016,roy_photoassociative_2016}) and \ce{LiBa} and \ce{LiCa}  in deeply bound states via conventional spectroscopy in references \citeText{dincan_electronic_1994} and \citeText{stein_spectroscopic_2013}.

\begin{figure}
		\includegraphics[width = \columnwidth]{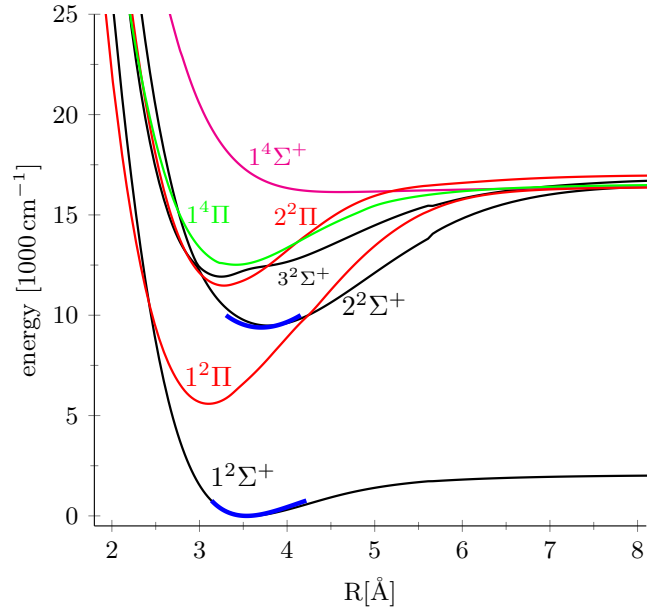}
		\caption{\label{fig:PotentialCurves} Potential curves of \ce{LiSr} from ab initio calculations\cite{gopakumar_ab_2013}. Thick curves represent RKR PECs from this work.}
	\end{figure}

	For many molecules of this class ab initio studies for potential energy curves (PECs), transition properties and electric dipole moments of the molecules exist, e.g. references \citeText{guerout_ground_2010,augustovicova_ab_2012,gopakumar_dipole_2014,pototschnig_vibronic_2017,shao_ground_2017}.
	Additional ab initio calculations for \ce{LiSr} have been performed by other groups\cite{gopakumar_ab_2011,kotochigova_ab_2011,gopakumar_ab_2013}. 	\figref{fig:PotentialCurves} shows a part of the potential energy scheme of \ce{LiSr} for the atom pair asymptotes Li(2s) + Sr(5s$^2$) and Li(2p) + Sr(5s$^2$) which is almost degenerate with Li(2s) + Sr(5s5p $^3$P). The latter atom pair leads also to molecular quartet states.
	
	In this work, we present the first spectroscopic observation of \LiSr and its analysis from which the bottoms of the potential energy curves of the ground state \Xstate and the first excited state \bstate of \LiSr are derived. This was achieved by creating the molecular gas in a heatpipe oven and observing the thermal emission spectrum in the near infrared region, which was expected from the ab initio results shown in \figref{fig:PotentialCurves}. The spectrum gave no indication of more than one isotopologue. Therefore, \ce{^7Li^{88}Sr}, formed by the most abundant isotopes of \ce{Li} and \ce{Sr}, was most likely observed. Laser excitations of the molecule were performed and were essential for an unambiguous assignment of the dense spectrum.
	For the observed $2 ^2\Sigma^+ \leftrightarrow 1 ^2\Sigma^+$ system, molecular parameters are derived. \figref{fig:PotentialCurves} shows an avoided crossing between the excited states \astate and $2^2\Sigma^+$. Thus perturbations are expected in the spectrum and  estimations for the \astate state will be derived from the observed coupling to the \bstate state. A comparison of  results from ab initio studies with experimental findings of this work is presented.

\section{Experimental scheme}

	A sample of \ce{LiSr} was prepared in a heatpipe and its thermal emission spectrum recorded with a Fourier Transform spectrometer (FTS). In a second experimental step, selected transitions of the molecule were excited by  a tunable external cavity diode laser and the resulting fluorescence was resolved through the FTS.

	A heatpipe of \SI{88}{\cm} in length and \SI{3}{\cm} in diameter was filled with about \SI{25}{\g} of \ce{Sr} and about \SI{2}{\g} of \ce{Li}. As a buffer gas, \SI{30}{\milli\bar} of argon was used.  A mesh was installed in the heatpipe to enable reflow of material condensing in the cold areas to the ends. Both ends were closed with BK7 windows tilted by few degrees against the optical axis. One end was used for imaging the emitted light to the spectrometer, the other end opened to a beam dump for the laser beam. A \SI{40}{\cm} long region in the center of the heatpipe was heated to \SI{915}{\celsius} while the ends were kept at room temperature. The thermal emission spectrum of \ce{LiSr} appears at about \SI{870}{\celsius}. 
	
	The spectrum was recorded with a Bruker FTS (IFS 120HR). Beam splitters for the near infrared  and an IR-enhanced Silicon avalanche photodiode (S11519-30, Hamamatsu) were installed. An optical low-pass filter (FGL850S, Thorlabs GmbH) and electronic filters in the detection circuit were applied to restrict the spectrum to the region of interest. It should be noted that the response of the detector vanishes around \SI{8500}{\kay}, which conveniently suppresses the detection of the rise of the thermal emission for a temperature around \SI{900}{\celsius}.

	\begin{figure}
		\includegraphics[width = \columnwidth]{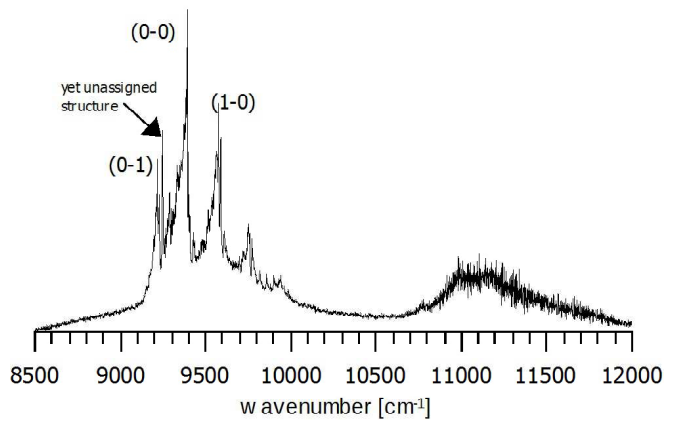}
		\caption{\label{fig:emSpectrum} Thermal emission spectrum of \ce{LiSr} at \SI{915}{\celsius} with a resolution of \SI{1}{\kay}.}
	\end{figure}

	The recorded spectrum ranges from \SI{8000}{\kay} to \SI{12500}{\kay}, an example is given  in \figref{fig:emSpectrum}. In the range from \numrange{9000}{10000} \si{\kay}, \ce{LiSr} dominates the spectral structure. Three prominent bands can be seen from \SI{9100}{\kay} to \SI{9600}{\kay}, tentatively assigned to the \bands{$v^{\prime}$}{$v^{\prime\prime}$}=\bands{0}{0}, \bands{0}{1} and \bands{1}{0} bands. Beyond the \bands{1}{0} band there are weaker structures seemingly following a band pattern. They can not be assigned so far. Starting around \SI{10500}{\kay}, a spectrum without obvious band structure can be seen. This structure coincides roughly with the expected $3^2\Sigma^{+} \leftrightarrow 1^2\Sigma^{+}$ electronic system from the ab inito calculations \cite{gopakumar_ab_2011}, but it is overlapped by the \ce{Li_2} spectrum.
	
	\begin{figure}
		\includegraphics[width = \columnwidth]{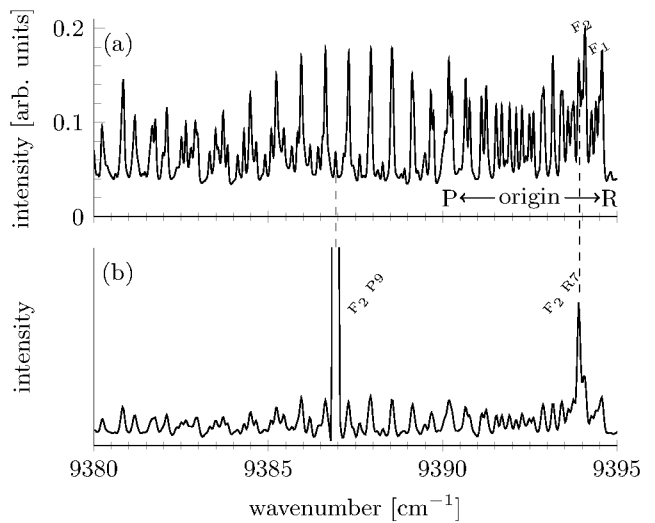}
		\caption{\label{fig:rotSpectrum} Rotational structure near the \bands{0}{0} bandhead at \SI{915}{\celsius} with a resolution of \SI{0.03}{\kay}(a) and superimposed with laser light at \SI{9386.932}{\kay} and a fluorescence line with a resolution of \SI{0.05}{\kay} (b). The \fone \ and \ftwo \ lines make two distinct bandheads.}
	\end{figure}

	\figref{fig:rotSpectrum} (a) shows a part of the recorded emission spectrum of \ce{LiSr} in detail. To resolve the rotational structure, the thermal emission spectrum was recorded with \SI{0.03}{\kay}. It is given in the supplementary material. The Doppler width is expected to be \SI{0.023}{\kay} for \LiSr at \SI{915}{\celsius} which justifies the selected resolution of the FTS. After averaging 1080 scans, a signal-to-noise ratio of about \num{250} can be achieved.
	
	In \figref{fig:rotSpectrum} (a) the band heads for \fone \ and \ftwo \  of the \bands{0}{0} band can be seen on the right side. The R branch turns at $N^{\prime\prime} \approx 12$. Most of the higher peaks are due to more than one transition line. The decreasing intensity towards \SI{9380}{\kay} is due to a perturbation in the R branch. The same perturbation is reflected in the P branch around \SI{9350}{\kay}. See sections \ref{sec:lineAss} and \ref{sec:pert} for details on the perturbation. Starting around \SI{9388}{\kay} towards lower wavenumber, the R branch of the \bands{1}{1} band starts to become visible from the background of the \bands{0}{0} emission.
	
	To observe laser induced fluorescence (LIF), the gas was excited with laser powers up to \SI{100}{\mW}. A diode laser in a Littrow configuration stabilized by a wavemeter (WS-U, High Finesse GmbH) was used to access the range from \SI{9200}{\kay} to \SI{10600}{\kay} with an accuracy of \SI{20}{\MHz}. The beam was collimated to a diameter of \SI{2}{\mm} and aligned with the heatpipe axis. LIF in the center of the heatpipe was imaged into the FTS minimizing stray light in the detection as the filters could not sufficiently block the backscattered laser light. The LIF spectra were recorded with a resolution of \SI{0.05}{\kay} and an example is shown in \figref{fig:rotSpectrum}~(b).  The fluorescence line is greatly enhanced compared to the thermal emission lines and has a signal-to-noise ratio of ca. \num{100} by averaging only 10 scans. In the example, the corresponding emission line in the pure thermal spectrum is an overlap of the lines \fone \ R18, \ftwo \ R7 and \ftwo \ R14. The LIF spectrum relates undoubtedly P or R lines for an excited rotational level. This is very important for the unambiguous assignment given in section \ref{sec:lineAss}.
	
	The observed LIF shows PR-doublets in bands \bands{$v^{\prime}$}{$(v^{\prime\prime}\pm 1)$} directly neighbouring the excited band \bands{$v^{\prime}$}{$v^{\prime\prime}$}. Rotational satellites from collisional relaxation were sometimes recorded but never spanned more than about five rotational levels. Long vibrational progressions were not observed as expected from the PECs in \figref{fig:PotentialCurves}. Laser excitations in the structure  from \SI{10500}{\kay} to \SI{12000}{\kay} have so far revealed PR-doublets that could be attributed to \ce{Li2}\cite{coxon_application_2006}.
	
	The line position was determined by fitting one or multiple Gaussian curves to a spectral line. The average frequency uncertainty given by the fit is close to the Doppler width. Where this method was not successful due to too many overlapping lines, the center of the spectral line was used as frequency and the FWHM was used as uncertainty. In the further analysis, the uncertainty was adjusted to be at least \SI{0.02}{\kay}.

\section{Line assignment}
\label{sec:lineAss}
	Based on the ab initio calculations from reference \citeText{gopakumar_ab_2013}, the observed band structure can be expected to be composed of $2^2\Sigma^{+} \leftrightarrow 1^2\Sigma^{+}$ transitions (see \figref{fig:PotentialCurves}). LIF experiments with the laser tuned to lines of the most intense band showed associated P and R lines only in the next visible band of lower frequency. Therefore, these bands can be tentatively assigned to be the \bands{0}{0} and \bands{0}{1} bands of this electronic system. 
	
	A $^2\Sigma^{+}$ state can be adequately described in Hund's coupling case (b) with basis vector $\ket{\Lambda, (N,S)J}$, where $\Lambda$ is the quantum number of the projection on the molecular axis of the orbital angular momentum (here zero), $\hat{N}$ is the total angular momentum without spins and $\hat{S}$ is the electron spin.  $\hat{J} = \hat{N} + \hat{S}$ is the total angular momentum of the molecule excluding nuclear spins. The energies of the rovibrational states can be expressed with the conventional Dunham expansion \cite{herzberg_spectra_1950}
	
	\begin{equation}
	E(v,N) = \sum_{m,n} \mathrm{Y}_{m,n}(v+\nicefrac{1}{2})^m[N(N+1)]^n.
	\label{eq:DunhamExpansion}
	\end{equation}
	
	For a doublet state, levels with $J=N+1/2$ or $J=N-1/2$ are labeled by \fone\ or \ftwo, respectively. The energy differences between the \fone \ and \ftwo \ components are then attributed to the spin-rotation coupling, given by the Hamiltonian $\gamma\mathbf{\hat{S}}\cdot\mathbf{\hat{N}}$\cite{herzberg_spectra_1950}, which is added to the rovibrational energies.  $\gamma$ is the coupling constant. The energy of a $J$ level thus evaluates to
	\begin{subequations}
	\begin{align}
		E_1(v,J)&= E(v,N) + \nicefrac{\gamma}{2} \times N  &\quad\mathrm{for\; F}_1 \\
		E_2(v,J)&= E(v,N) - \nicefrac{\gamma}{2} \times (N+1) &\quad\mathrm{for\; F}_2
	\end{align}
	\label{eq:SRcoupling}
	\end{subequations}
	for the vibrational and rotational quantum numbers $v$ and $N$ and $E(v,N)$ given by the Dunham expression. The strength of the spin-rotation coupling can change with the internuclear separation and thus a slight dependence on $v$ and $N$ was observed. For mnemonic reasons, this dependence is modeled in analogy to the Dunham expansion:

	\begin{equation}
		\gamma(v,N) = \sum_{m,n} \gamma_{m,n}(v+\nicefrac{1}{2})^m[N(N+1)]^n.
		\label{eq:gammaExpansion}
	\end{equation}
	
\begin{figure}
		\includegraphics[width = \columnwidth]{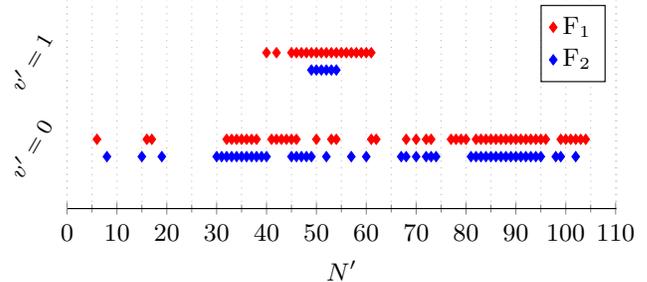}
		\caption{ Rovibrational levels of the \bstate state which were addressed with a laser excitation from $v^{\prime\prime} = 0,1,2$ in the \Xstate state}
       \label{fig:assignedLines}
	\end{figure}

	Many lines of the \bands{0}{0} band could be assigned to rotational transitions using frequency differences from fluorescence PR-doublets near the band head and the value for the rotational constant given in reference \citeText{gopakumar_ab_2013} as a first approximation.
	The corresponding fluorescence lines in the \bands{0}{1} band were assigned accordingly. Lines of both spin components \fone \ and \ftwo \ were assigned up to $N^{\prime\prime} =104$. \figref{fig:assignedLines} summarizes the levels of the state \bstate that were addressed in the LIF experiments. Because the experimental procedure gives no information about the sign of the spin-rotation constant $\gamma$,  a definite assignment of spectral lines to \fone \ or \ftwo \ could not be made. Inverting \fone \ and \ftwo \ will give the same result with opposite sign of $\gamma$.

\begin{figure}
		\includegraphics[width = \columnwidth]{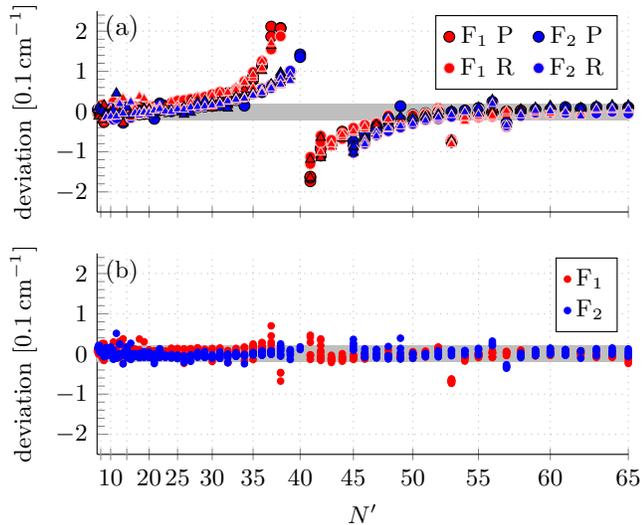}		
		\caption{ (a) Deviation of actual transition frequencies in the \bands{0}{0,1} bands from frequencies calculated with the Dunham model. Circles represent transitions in the \bands{0}{0} band and triangles represent transitions in the \bands{0}{1} band. (b) Deviations considering coupling between electronic states.  The grey area depicts the experimental  uncertainty. }
		\label{fig:deviation}
	\end{figure}

	While the ground state levels obtained  from the fluorescence progressions could be described with the formulas \eqref{eq:SRcoupling} and \eqref{eq:gammaExpansion}, the Dunham model proved insufficient to describe the energies of some rotational levels in the $v^\prime = 0$ manifold. These levels show  systematic deviations from the energies given by the Dunham series as displayed in \figref{fig:deviation}, suggesting a perturbation. For a range of about 20 rotational levels, centered around $N^\prime \approx 40$, a model including the spin-orbit coupling to \astate is developed (section \ref{sec:pert} below). For $N^\prime>68$, the perturbations become complicated and this part of the rotational manifold was therefore not taken into account for this work.

\begin{figure}
		\includegraphics[width = \columnwidth]{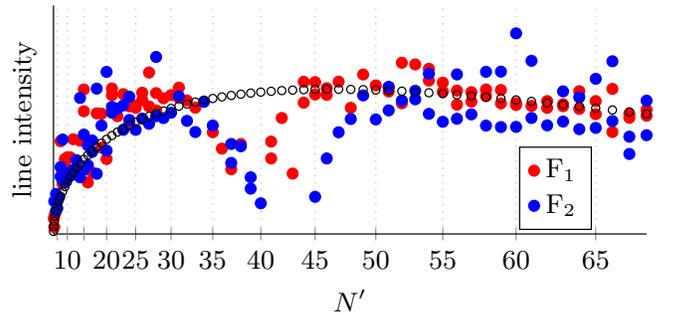}		
		\caption{ Intensities of thermal emission lines of the \bands{0}{0} band. Line intensities were always determined on a local scale subtracting the background. Overlapping lines were left out if their respective intensities could not be determined. Thermal populations calculated with $E(v^\prime=0,N^\prime) \approx \mathrm{Y}^\prime_{01}~\times~N^\prime(N^\prime+1)$ are shown for comparison (black circles). }
		\label{fig:Intensities}
	\end{figure}

	Another feature of the perturbation is that the lines with perturbed energy levels have reduced intensity. This was seen in the thermal emission and LIF spectra. \figref{fig:Intensities} shows the intensities from the emission spectrum for the perturbed range of quantum numbers identified above. For values of $N$ around 40 with large perturbation, the intensities reduce significantly and result in reduced or vanishing fluorescence from the most perturbed levels. For this reason, the lines from perturbing states, the so called extra lines, could not be identified because the LIF experiments yielded no identifiable response. This observation is in agreement with the expectation that the perturbation comes from the coupling to the \astate state, which has a low electronic transition moment to the ground state and unfavorable Franck-Condon factors, according to ab initio calculations \cite{gopakumar_ab_2013} (see also \figref{fig:PotentialCurves}).

	Also by LIF, a system of three connected bands was discovered which are identified as the \bands{1}{0}, \bands{1}{1} and \bands{1}{2} bands inserting the vibrational spacing of the ground state obtained from the fluorescence progressions. LIF experiments in the less intense spectral structure seen between the \bands{1}{0} band and \SI{10000}{\kay} in \figref{fig:emSpectrum} were unsuccessful, possibly due to insufficient laser intensity. For the \bands{1}{0} band, lines in the range of $N^{\prime\prime} = $ \numrange{40}{60} were assigned using the already derived rotational energies of the ground state (see upper part in \figref{fig:assignedLines}) but the excited state could not be described by the Dunham model. For this reason, only the $v'=0$ state with rotational levels up to $N^\prime=68$ is considered in this work.
	
	468 measured transition frequencies and 821 frequency differences of the PR-doublets along with their assigned quantum numbers were used for a linear fit of the Dunham coefficients for both $^2\Sigma^+$ states as given in \eqref{eq:DunhamExpansion} and  spin-rotation parameters  in \eqref{eq:gammaExpansion}. The transitions associated with an upper level that was recognized as perturbed were excluded from this fit.

\section{Perturbation}
\label{sec:pert}
	
	A substantial number of observed transition lines suggest perturbations in the \bstate state (see \figref{fig:deviation} (a)). For the lines with significant deviation from the Dunham model, the LIF experiments proved highly important in ascertaining the quantum number assignment since the PR-differences are governed by the involved (unperturbed) \Xstate levels, exclusively.
	
	According to reference \citeText{gopakumar_ab_2013}, the $1^2\Pi_{1/2}$ and $1^2\Pi_{3/2}$ states are energetically closest to the state under study.  $^2\Sigma^+$ and $^2\Pi$ states are coupled by the spin-orbit and rotational interaction.  Following reference \citeText{lefebvre-brion_perturbations_1986}, the matrix representation of the Hamiltonian in Hund's coupling case (a) with state vector $\ket{\Lambda,S,\Sigma,J}$ for a total angular momentum $J$ is derived, where $\Sigma$ is the quantum number of the projection of $\hat{S}$  and $\Omega = \Lambda + \Sigma$ is the quantum number of the projection of $\hat{J}$ to the molecular axis.  Hyperfine interaction has not been considered. No effects of the hyperfine structure by additional splitting or on  line shapes were observed and hence it can be assumed that its effects on line positions are negligible. The matrix is given in \tabref{tab:intMatrix}. The upper and lower signs correspond to the \fone \ and \ftwo \ states of a given total angular momentum $J$.

	\begin{table*}\caption{\label{tab:intMatrix} Interaction between the  $2^2\Sigma^{+}$ and  $1^2\Pi$ states in Hund's coupling case (a) for a set of vibrational levels $v_\Sigma, v_\Pi$. $\mathrm{E}_\mathrm{Dun} $ is the energy calculated from Dunham parameters, $\gamma$ is the spin-rotation constant, $A$ is the spin-orbit coupling constant, $B_v$ is the rotational constant for a given vibrational state. $T_\Pi$ is the electronic energy and $G_\Pi(v)$ the vibrational energy of $1^2\Pi$. The factor $p$ represents the expectation value $\braket{v_\Sigma |\mathbf{\hat{L}^{\pm}} |v_\Pi}$. Subscripts on the constants indicate a value for the $\Sigma$ or $\Pi$ state or a mixture thereof. Upper signs are for \fone and lower signs for \ftwo . }
		{%\footnotesize
			\centering
			\begin{tabular}{l|*3{c}} 
				& $\ket{^2\Sigma^{+}_{1/2}}$ & $\ket{^2\Pi_{1/2}}$ & $\ket{^2\Pi_{3/2}}$  \\
				\hline
				$\bra{^2\Sigma^{+}_{1/2}}$ & \begin{tabular}{@{}c} $\mathrm{E}_\mathrm{Dun}(v,N=J\mp \nicefrac{1}{2}) $ \\$ -\nicefrac{\gamma_\Sigma}{2} \times \left [1 \mp (J+\nicefrac{1}{2}) \right ] $ \end{tabular} & \begin{tabular}{@{}c} $\nicefrac{p}{2} \times \big[A_{\Sigma\Pi} - \gamma_{\Sigma\Pi} \, + $ \\ $2 \, B_{\Sigma\Pi} ( 1 \mp [J+\nicefrac{1}{2}] ) \big] $ \end{tabular}& \begin{tabular}{@{}c}$-p \times B_{\Sigma\Pi} \times $\\ $ \sqrt{J(J+1) -\nicefrac{3}{4}  }$\end{tabular}\\[10pt]
				$\bra{^2\Pi_{1/2}}$ & \begin{tabular}{@{}c} $\nicefrac{p}{2} \times\big[A_{\Sigma\Pi} - \gamma_{\Sigma\Pi} \, + $ \\ $2\, B_{\Sigma\Pi} ( 1 \mp [J+\nicefrac{1}{2}] ) \big] $ \end{tabular}  & \begin{tabular}{@{}c} $\mathrm{T}_{\Pi}+B_{\Pi,v} [J(J+1)+\nicefrac{1}{4}]+ $ \\ $G_\Pi(v)- (A_\Pi + \gamma_\Pi)/2$ \end{tabular} & \begin{tabular}{@{}c} $ (\nicefrac{\gamma_\Pi}{2} - B_\Pi ) \, \times $\\ $\sqrt{J(J+1) -\nicefrac{3}{4}  }$ \end{tabular} \\[10pt]
				$\bra{^2\Pi_{3/2}}$ & \begin{tabular}{@{}c}$-p \times B_{\Sigma\Pi} \times $\\ $ \sqrt{J(J+1) -\nicefrac{3}{4}  }$\end{tabular}  & \begin{tabular}{@{}c} $ (\nicefrac{\gamma_\Pi}{2} - B_\Pi ) \, \times $ \\ $ \sqrt{J(J+1) -\nicefrac{3}{4}  }$ \end{tabular} &  \begin{tabular}{@{}c} $\mathrm{T}_{\Pi}+B_{\Pi,v} [J(J+1)-\nicefrac{7}{4}]+ $ \\ $G_\Pi(v) + (A_\Pi - \gamma_\Pi)/2$ \end{tabular}
			\end{tabular}
		}
	\end{table*}

	The diagonal entries describe the energies of the $^2\Sigma^+$ and $^2\Pi$ states without coupling. $\mathrm{E}_\mathrm{Dun}$ for the \bstate state is the rovibrational energy according to \equref{eq:DunhamExpansion}. For the $^2\Pi$ state, the energy levels are given by the electronic energy $\mathrm{T}_\Pi$, vibrational energy $G_\Pi(v)$ and rotational constant for a vibrational level $B_{\Pi,v}$. Three additional parameters appear in the matrix: the coupling constants of the spin-rotation interaction and the spin-orbit interaction, $\gamma$ and $A$, and the factor $p = \braket{v_\Sigma |\mathbf{\hat{L}^{\pm}} |v_\Pi}$, which will be approximated as the product of an overlap integral $\braket{v_\Sigma |v_\Pi}$ of the coupled vibrational states and the expectation value $\braket{\Pi |\mathbf{\hat{L}^{+}}| \Sigma}$over the electronic space. Assuming the electronic states belong to $L = 1$, $\braket{\mathbf{\hat{L}^{\pm}} }$ evaluates to $\sqrt{2}$. For all parameters, a subscript indicates the corresponding electronic states. The non-diagonal terms for $\Delta \Omega =0$ come from spin-orbit interaction and those for $\Delta \Omega =\pm1$ are rotational interactions, the later ones also couple $^2\Sigma^+_{-1/2}$ and $^2\Pi_{+1/2}$.
	
	To keep the number of fit parameters low, some simplifications had to be made because we only have data for state $2^2\Sigma^+$, which couples strongly through spin-orbit interaction to the component $\Omega=1/2$ of \astate but only weakly by rotation to $\Omega=3/2$. Thus we reduce the $3\times3$-matrix to the $2\times2$ case. The coupling constants $A_{\Sigma\Pi}$ and $\gamma_{\Sigma\Pi}$ cannot be separate in fitting experimental data. They are combined into the constant $d_{\Sigma\Pi}$. For the same reason we combine  $A_\Pi$ and $\gamma_\Pi$ to the effective constant A. From the difference of the $1^2\Pi_{1/2}$ and $1^2\Pi_{3/2}$ PECs given in the supplement of reference \citeText{gopakumar_ab_2013}, $A_\Pi$ can be estimated to be \SI{118}{\kay}.  The approximate interaction matrix is shown in \tabref{tab:redIntMatrix}.

	\begin{table}\caption{\label{tab:redIntMatrix} Reduced interaction matrix between the  $2^2\Sigma^{+}$ and  $2^2\Pi_{1/2}$ state. $d_{\Sigma\Pi} = A_{\Sigma\Pi} - \gamma_{\Sigma\Pi}$ and $A = A_\Pi \pm \gamma_{\Pi} \approx A_\Pi$. The $^2\Pi_{3/2}$ state is ignored.}
		{%\footnotesize
			\centering
			\begin{tabular}{l|*2{c}} 
				& $\ket{^2\Sigma^{+}_{1/2}}$ & $\ket{^2\Pi_{1/2}}$  \\	
				\hline
				$\bra{^2\Sigma^{+}_{1/2}}$ & \begin{tabular}{@{}c} $\mathrm{E}_\mathrm{Dun}(v,N=J\mp \nicefrac{1}{2}) $ \\$ -\nicefrac{\gamma_\Sigma}{2} \times \left [1 \mp (J+\nicefrac{1}{2}) \right ] $ \end{tabular} & \begin{tabular}{@{}c} $\nicefrac{p}{2} \big[d_{\Sigma\Pi} \, + $ \\ $2\, B_{\Sigma\Pi} ( 1 \mp [J+\nicefrac{1}{2}] ) \big] $ \end{tabular}\\[10pt]
				$\bra{^2\Pi_{1/2}}$ & \begin{tabular}{@{}c} $\nicefrac{p}{2} \big[d_{\Sigma\Pi} \, + $ \\ $2\, B_{\Sigma\Pi} ( 1 \mp [J+\nicefrac{1}{2}] ) \big] $ \end{tabular}  & \begin{tabular}{@{}c} $\mathrm{T}_{\Pi} -\nicefrac{A}{2} + G_\Pi(v)    $ \\ $+B_{\Pi,v} [J(J+1)-1]$ \end{tabular}
			\end{tabular}
		}
	\end{table}

	\begin{figure}
		\includegraphics[width = \columnwidth]{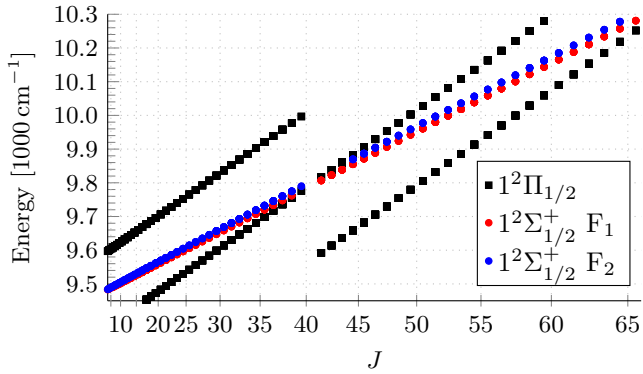}		
		\caption{ Rotational energies of the $v^\prime = 0$ level of the $2^2\Sigma^{+}$ state and the three closest vibrational levels of the $1^2\Pi_{1/2}$  state.}
       \label{fig:energyLadders}
	\end{figure}
	
	To characterize the perturbation seen in \figref{fig:deviation} (a), knowledge about the crossing of the rotational states of the \bstate and  \astate states and the various coupling strengths are required. For the \astate state we start with parameters taken from reference \citeText{gopakumar_ab_2013}. In order to come close to the observed resonant perturbation, $T_\Pi$ and $B_\Pi$ were adjusted to move the crossing points of the rotational ladders of one vibrational level of $1^2\Pi_{1/2}$ and $v^\prime_{\Sigma}= 0$ into the range of maximal deviation (see \figref{fig:energyLadders}). The variation of the sign of the deviation shows that the rotational constant of the perturbing state must be larger than that of state \bstate to obtain  repelling levels in the observed direction.
	Taking this initial choice of parameters for the $1^2\Pi_{1/2}$ state, only $p$, $d_{\Sigma\Pi}$ and $B_{\Sigma\Pi}$ are unknown for a fit.

	Since the \astate state is not known well enough to assign $v_\Pi$ unambiguously and thus  to calculate the desired overlap integral with the \bstate state with satisfactory reliability, we incorporate the parameter $p$ into $d_{\Sigma\Pi}$ and set $B_{\Sigma\Pi}$ initially to zero.

	\begin{table*}
		\caption{\label{tab:DunPar}Dunham and spin-rotation parameters for the first two $^2\Sigma^{+}_{1/2}$ states and the first  $^2\Pi_{1/2}$ state of \LiSr. The parameters give an accurate description for levels with $N<105, v= 0,1$ and $40 \leq N \leq 60, v=2$ in the \Xstate state, $N < 69, v=0$ in the \bstate state and for $N<69, v \approx 15$ in the \astate state.  All values given in \si{\kay}.}

		\begin{ruledtabular}
			\begin{tabular}{*{5}lr}
				\noalign{\vskip 5pt} 
				\multicolumn{6}{c}{$1^2\Sigma^{+}_{1/2}$} \\
				$\mathrm{Y}_{0n}$ & $\mathrm{Y}_{1n}$ & $\mathrm{Y}_{2n}$& $\gamma_{0n}$ & $\gamma_{1n}$ & $n$ \\
				\num{0.0} & \num{182.9305 +- 0.0054} & \num{-3.0263 +-  0.0026}& \num{8.88 +-  0.46 e-3} & \num{-5.28 +-  0.16 e-4} & 0\\
				\num{2.072284 +-   0.000079 e-1} & \num{-3.3395 +-  0.0023 e-3} & \num{-1.0525 +-  0.0092 e-4} &-&-& 1\\
				\num{-1.0297 +-  0.0017 e-6} & \num{-3.662 +-  0.042 e-8} & - &-&- & 2\\
				\num{ -5.89 +-  0.10 e-12} & \num{-2.187 +-  0.032 e-12} & - &-&- &3\\
				\hline
				\noalign{\vskip 5pt} 
				\multicolumn{6}{c}{$2^2\Sigma^{+}_{1/2}$} \\
				\num{9389.2125 +-  0.0026} & \num{186.94} \footnote{Value from reference \citeText{gopakumar_ab_2013}} &- &\num{4.651 +-  0.046 e-2} &- & 0 \\
				\num{1.890807 +-  0.000081 e-1} & - &- & - &- & 1 \\
				\num{-7.922  +- 0.018 e-7} & - & -&-&- & 2 \\
				\num{3.77  +- 0.14 e-12} & - &- &-&- &3 \\
				\hline
				\noalign{\vskip 5pt} 
				\multicolumn{6}{c}{$1^2\Pi_{1/2}$} \\
				\num{5403.7 +-  2.4}\footnote{Disregarding an offset of $A/2$} & \num{285.634} $^{\text{a}}$ & \num{-1.91289} $^{\text{a}}$ &- &- & 0 \\
				\num{2.837 +-  0.014 e-1} & \num{-2.01 e-3} $^{\text{a}}$& - &- &- & 1\\
				\hline
				\noalign{\vskip 5pt} 
				\multicolumn{4}{r}{$p/2 \times d_{\Sigma\Pi}$:} & \multicolumn{1}{l}{\num{2.644 +- 0.043} } & 
			\end{tabular}
			
		\end{ruledtabular}

	\end{table*}

	The rovibronic parameters $T_\Pi$ and $B_\Pi$ and $d_{\Sigma\Pi}$ were varied in order to minimize the deviation for all data points shown in \figref{fig:deviation} by a non-linear least-squares fit using the energies obtained after matrix diagonalization.
	From this fit, unperturbed levels of the \bstate state and corresponding transition frequencies were constructed and then applied in the linear fit with energies represented by \equref{eq:SRcoupling} for improved Dunham coefficients for the \Xstate and \bstate states. For this, 1534 observations (transition frequencies from the thermal emission spectrum, fluorescence lines along with frequency differences from LIF spectra) were used. Such procedure was cycled and after three iterations the Dunham coefficients changed by less than the estimated standard deviation from the linear fit. Thus we obtained convergence of the iterative fitting procedure. Finally, the stability of this solution was checked by fitting the parameters for the \bstate and $1^2\Pi^+_{1/2}$ states along with $d_{\Sigma\Pi}$ simultaneously in the non-linear fit step. Additionally, we allowed the variation of the J-dependent off-diagonal term by $B_{\Sigma\Pi}$. It turned out that this contribution is insignificant in the range of observations.
	The final coefficients are listed in \tabref{tab:DunPar} together with estimated standard deviations from the linear fit. The standard deviation of the fit was 0.492.
		
	\figref{fig:deviation} (b) shows that the perturbation is well described because most deviations lie within the gray area, which represents the experimental uncertainty. For the \fone \ states closest to the perturbation, no ideal description could be achieved. For improving the modeling we would like to have data of the $1^2\Pi$, especially the observation of the extralines expected around the perturbed lines of $2^2\Sigma^+$, but we were so far unsuccessful in our efforts.
	
	Additionally, a local deviation around $N^\prime = 53$ for \fone \ and  $N^\prime = 57$ for \ftwo can be seen in \figref{fig:deviation}. These might indicate a crossing of \bstate with $1^2\Pi_{3/2}$. That state was ignored in the simplified interaction model and therefore the aforementioned lines were ignored in the fitting process. A manual adjustment of the spin-orbit coupling parameter $A_\Pi$ to shift the $1^2\Pi_{3/2}$ state to the corresponding energies gives a value of $A_\Pi = $\SI{88 +- 2}{\kay}, which is still close to the ab initio value, but we believe that this single observation is not yet conclusive.

\section{results and discussion}
	The deeply bound rovibrational levels of the \Xstate and \bstate states of \LiSr were modeled using the thermal emission and LIF spectra. 
	The $v^{\prime\prime} =0,1$ levels of the ground state \Xstate could be described up to $N^{\prime\prime} = 105$ and the $v^{\prime\prime} = 2$ level with $N^{\prime\prime}$ ranging from \numrange{41}{64} by the Dunham parameters including spin-rotation. 
	Transitions in this system were found to take place mainly between states with the same or a directly neighbouring vibrational quantum number. This restricts the study of the molecule via LIF experiments and becomes a time consuming work because only short progressions were observed. To investigate higher vibrational states for a more complete ground state potential, other studies including higher lying electronic states need to be employed.
	
	With the perturbation model from section \ref{sec:pert}, molecular parameters for the $v^\prime = 0$ level of the \bstate state with $N^\prime < 69$ were derived. Including higher rotational states in the evaluation was not yet successful due to the complex perturbation structure. The isolated perturbation of the rovibrational levels for $N^\prime<69$ gave an opportunity to gauge the strength of the \astate -- \bstate coupling and gain first insights to the \astate state. Since the \astate state could not be observed directly with the applied method, this knowledge will be the initial ingredient when incorporating transition lines of higher \bstate levels into the analysis of the spectrum. 
	
	The derived molecular parameters are given in \tabref{tab:DunPar}. Since only a finite number of parameters were fitted, the given parameters are not the true Dunham parameters; they are significantly affected by the truncation of the power expansion. For example, $\mathrm{Y}_{01}$ of \bstate is the rotational constant $B_0$ of the evaluated vibrational level. Utilizing the observed perturbation, a  parametrization of the energy levels of the $1^2\Pi_{1/2}$ state in the neighbourhood of $v_\Sigma =0$ of the \bstate state was possible. The effective coupling parameter, $p/2 \times d_{\Sigma\Pi} = \SI{2.644}{\kay}$ is fairly low compared to the estimated spin-orbit parameter around \SI{100}{\kay}, indicating the weak overlap of the involved vibrational levels of the states \astate and $2^2\Sigma^+$. A fit with the $J$-dependent coupling parameter $B_{\Sigma\Pi}$ (see \tabref{tab:redIntMatrix}) gave no significant improvement and shows variations in the value of $\gamma_{00}$ of \bstate of less than one percent. Thus the effective spin-rotation interaction of \bstate does not originate from the investigated local perturbation but will be the integrated effect of the full manifold of \astate -- \bstate coupling and/or couplings to other $^2\Pi$ states. 
	For \LiSr it is possible to observe the \astate state indirectly due to the long $N$ interval of interaction with the \bstate state. This is not guaranteed for all molecules and has not yet been observed in \ce{LiCa}\cite{ivanova_x_2011, stein_spectroscopic_2013}.

	\begin{table*}
		\caption{\label{tab:constComp}Comparison of measured spectroscopic constants of \LiSr with results of various ab initio works. All values given in \si{\kay}, except $R_e$ which is given in \AA. }
		\begin{ruledtabular}
			\begin{tabular}{llccccccr}
				&Method&$R_e$ &$D_e$ &$\omega_e \approx \mathrm{Y}_{10}$&$\omega_e x_e \approx -\mathrm{Y}_{20}$ &$B_e \approx \mathrm{Y}_{01}$ &$T_e $ & Ref.\\
				\hline
				\noalign{\vskip 5pt} 
				$1^2\Sigma^{+}_{1/2}$&UCCSD(T) & 3.55\phantom{0} & 2367\phantom{00} & 182.2\phantom{0} & - & -& 0 & \citeText{kotochigova_ab_2011}$\phantom{^{\text{a}}}$\\
				&CCSD(T) & 3.531 & 2226.4\phantom{0} & 182.1\phantom{0} & 4.29 & 0.203$\phantom{00^{\text{a}}}$ & 0 & \citeText{gopakumar_ab_2011}$\phantom{^{\text{a}}}$\\
				&SO-MS-CASPT2 & 3.579 & 2075.26 & 168.62 & - & 0.2036$\phantom{0^{\text{a}}}$ & 0 & \citeText{gopakumar_ab_2013}\footnote{Values have been converted to \LiSr.}\\
				&MRCI & 3.574 & 2483\phantom{.00} & 179.1\phantom{0} & 3.22 & - & 0 & \citeText{pototschnig_vibronic_2017}$\phantom{^{\text{a}}}$\\
				&spectroscopy & \phantom{$^\text{a}$}3.545\footnote{From RKR calculation} & - & 182.93 & 3.03 & 0.207$\phantom{00^{\text{a}}}$ & 0 & this work$\phantom{^{\text{a}}}$\\
				\hline
				\noalign{\vskip 5pt}    
				$2^2\Sigma^{+}_{1/2}$&SO-MS-CASPT2 & 3.785 & 6860.54 & 186.94 & - & 0.1711$\phantom{0^{\text{a}}}$ & 9488.63 & \citeText{gopakumar_ab_2013}$^{\text{a}}$\\
				&MRCI & 3.728 & 7811\phantom{.00} & 183.0\phantom{0} & 1.07 & - & 9469\phantom{.00} & \citeText{pototschnig_vibronic_2017}$\phantom{^{\text{a}}}$\\
				&spectroscopy & $\phantom{^{\text{a,b}}}$3.712$^\text{b,}$\footnote{$R_0 = \SI{3.708}{\AA}$} & - & $\phantom{^{\text{a}}}(186.94)\footnote{RKR potentials were calculated with $\omega_e$ taken from reference \citeText{gopakumar_ab_2013}.}\phantom{0}$ & - & 0.18908\footnote{$B_0$, not $B_e$, because only one vibrational state was involved.} & $\phantom{^{\text{b,f}}}$9388.31$^\text{b,}$\footnote{$T_0=\SI{9482.683}{\kay}$}& this work$\phantom{^{\text{a}}}$\\
			\end{tabular}
		\end{ruledtabular}
	\end{table*}

	Using the Dunham parameters, PECs for the lowest vibrational states were calculated using the RKR-Method (see e.g. reference \citeText{telle_fcfrkr_1982} and references therein). They are indicated by thick, blue lines in \figref{fig:PotentialCurves} and show essential agreement with the ab initio PECs but energy displacements are in the order of \SI{100}{\kay} within the thick lines. Deriving Franck-Condon-factors\cite{herzberg_spectra_1950} from these PECs, we get the confirmation by the values why the emission spectrum of $2 ^2\Sigma^+ \leftrightarrow 1 ^2\Sigma^+$ is concentrated on few vibrational levels and why no long vibrational progressions are observed from laser excitations. The authors of reference \citeText{pototschnig_vibronic_2017} found this to be generally expected for alkali-alkaline earth dimers.

	\tabref{tab:constComp}  shows a comparison of some spectroscopic constants derived in this work with several ab initio calculations. Overall, the rotational constant $B_e\approx Y_{01}$, the vibrational constant $\omega_e\approx Y_{10}$ and the electronic energy $T_e$ were found to be smaller than the ab initio values. The value of $\omega_e$ varies by about \SI{10}{\percent} between different studies, two of which differ by less than \SI{1}{\percent} from the experimentally found value for the \Xstate state. The equilibrium internuclear distances along with $B_e$ values deviate by only a few percent. The values for $T_e$ of the \bstate state agree within \SI{100}{\kay}. The potential depth from the ab initio calculations disagree between each other by a few hundred \si{\kay} for the ground state but by up to \SI{1000}{\kay} for the \bstate state. An experimental result for this parameter for the $^2\Sigma^+$ states could not be achieved here. A comparison for the \astate state is not yet possible because of the single perturbation level, which would be about $v_\Pi=15$ applying the ab initio results from ref. \citeText{gopakumar_ab_2013}.
	
	Further work on \ce{LiSr} will be done to describe the transition lines with $N^\prime > 68$ in a more extensive model of the perturbation based on the findings presented here.

	Together with data from different publications for \ce{^7Li^{40}Ca}\cite{ivanova_x_2011, stein_spectroscopic_2013} and \ce{^7Li^{138}Ba} \cite{dincan_electronic_1994}, a trend in the molecular constants seems to emerge. For the $\Sigma$ states, the product of the reduced mass and the rotational constant decreases with increasing reduced mass while the product of the reduced mass and the spin-rotation coupling increases with the reduced mass. This finding relates nicely to the increase of the spin-orbit interaction from Ca via Sr to Ba.
	
	The spectrum of \ce{LiSr} is fairly dense and thus difficult to analyze due to the many overlapping lines. Thus alkali-alkaline earth dimers with a larger reduced mass should have denser spectra and the LIF method would be immensely advantageous to obtain simplified spectra to help in the assignment of quantum numbers. Work on \ce{KCa} is in progress in our lab and confirms this expectation.

\section*{supplementary material}

	See supplementary material for the full recorded thermal emission spectrum and a list of assigned emission lines for $N^\prime < 69$ of the \bands{0}{0} and \bands{0}{1} bands.

\section*{acknowledgments}
	This work received financial support from the Deutsche Forschungsgemeinschaft (DFG).

%\section{appendixes}

\section*{references}

\bibliography{LiSr_17}

\end{document}